\documentclass[aps,prb,showpacs,twocolumn]{revtex4-1}

\usepackage{amsmath,graphics}
\usepackage[next]{inputenc}
\usepackage[dvips]{epsfig}

\def\be{\begin{equation}}
\def\ee{\end{equation}}

\def\bi{\begin{itemize}}
\def\ei{\end{itemize}}
\def\bn{\begin{enumerate}}
\def\en{\end{enumerate}}
\def\bea{\begin{eqnarray}}
\def\eea{\end{eqnarray}}

\def\ba{\begin{array}}
\def\ea{\end{array}}
\def\bd{\begin{displaymath}}
\def\ed{\end{displaymath}}

\begin{document}
\title{Implementation of Single-qubit and CNOT Gates by Anyonic Excitations of Two-body Topological Color Code}
\author{Mehdi Kargarian}
\email[]{kargarian@physics.utexas.edu}
\affiliation{Department of Physics, The University of Texas at Austin, Austin, TX 78712, USA}

\begin{abstract}
The anyonic excitations of topological two-body color code model are used to implement a set of gates. Because of two-body interactions, the model can be simulated in optical lattices. The excitations have nontrivial mutual statistics, and are coupled to 
nontrivial gauge fields. The underlying lattice structure
provides various opportunities for encoding the states of a logical
qubit in anyonic states. The interactions make the
transition between different anyonic states, so being logical
operation in the computational bases of the encoded qubit. Two-qubit
gates can be performed in a topological way using the braiding of
anyons around each other.

\end{abstract}
\date{\today}

\maketitle


\section{Introduction \label{introduction}}
A set of universal quantum gates is a set of basic gates in which any
operation in a quantum computer can be decomposed. This means that
any unitary operation can be expressed as a finite sequence of the
gates from the set.\cite{nielsen, miguel:rmp02} The gates are used to process and
transform the encoded information on the quantum register. They are
unitary operations acting on one or two qubits and transform their
states. For instance the Hadamard gates and rotation about axes are
among single-qubit gates, while the controlled-Phase gates is a
two-qubit gate acting between control and target qubits. Any
arbitrary unitary operation acting on array of qubits can be
synthesized using above gates, i.e. they form a \emph{universal
set}.\cite{jean}

The main challenge of quantum computation is to design ways in which
the universal gates can be implemented avoiding the accumulation of
errors during the processing. Any information processing task must
be robust against decoherence, that is, as the gates are
implemented, the stored information are not read out.

Topological models provide an opportunity to secure information from
decoherence and perform implementation on encoded information
fault-tolerantly as well.\cite{mochon,freedman,nayak:rmp08} A form to 
achieve fault-tolerance is by means of self-correcting quantum computers.\cite{hector:arxiv09,hamma:arxiv09} In the topological models, the quantum information is encoded on the global degrees of
freedom. Since operating on the global degrees of freedom deserves
non-local operations acting on large number of qubits, the local
perturbations are not able to destroy the stored information. In
fact the ground states of the topological model can be used as
topological quantum memory.\cite{dennis} The storing of quantum
information is interesting, but one needs to process the stored
information. To perform computation on the stored information, one may use
the topological properties of the excitation appearing above the
ground state. The excitations have exotic statistics that are
neither fermion nor boson. They are anyons, instead. The non-trivial braiding
of anyons can be used to construct gates. If under the monodromy
operations on the excitation (winding of excitations around each other) the wavefunction of the system acquires
a global phase, the respective anyons are called abelian anyons. But
they could also be non-abelian in the sense that the evolution of the wave
function is described by unitary matrices.

The famous Kitaev model on a honeycomb lattice have abelian and non-abelian
anyonic states appearing at different regimes of
couplings.\cite{kitaev1} In particular, the
emerging abelian anyons can be used to perform single-qubit and
two-qubit gates.\cite{Lloyd, pachos} Although such implementation by
abelian anyons is interesting by itself, but creating a purely anyonic state with only low energy vortices in this model is challenging. This is important for physical
applications. Anyonic excitations are created by applying local spin operators
on the ground state of the Kitaev honeycomb model. But, any attempt
to create anyons is spoiled by creation of high energy
fermions.\cite{vidal1,vidal2}

Topological color code model (TCC)\cite{hector1,hector2} presents another topological
stabilizer model with enhanced computational capabilities such as transversal implementation of 
whole Clifford group. The code appears as ground state subspace of the
two-body Hamiltonian defined on a rubby lattice.\cite{kargarian1, kargarian2}
Emerging high energy fermions belong to different classes each of
one color charge. In each class high energy excitations have
fermionic statistics while fermions associated to different classes
have mutual semionic statistics.\cite{hector3} This latter point
implies that high-energy excitations are not only fermions but also anyons which is absent in 
the Kitaev model. 

The aim of this paper is to use the anyonic character of the
emerging high energy fermions in order to implement gates. Recently, experimental realizations of topological
error correction codes have been implemented,\cite{expriment1} as well as
experimental proposals for TCCs using Rydberg atoms
\cite{expriment2}. We use various
ways of encoding qubits. Once the qubit states are encoded
into anyonic fermions, the proper implementation of single-qubit and
two-qubit gates are instructed by manipulation of anyons. The fact that the quantum state
of hard-core bosons (anyonic fermions) realized in optical lattices enlightens and motivates our construction 
for encoding and manipulation of information.\cite{exp_hardcore} The creation and manipulation of anyonic fermions are done by interactions in the Hamiltonian which can be controlled in optical lattices. In particular, we will show that the manipulation of anyonic fermions can be used to implement 
the single-qubit $X$ and $Z$ gates in a non-topological way, and the two-qubit CNOT gate is topologically performed via the braiding of anyonic fermions. 
 
The paper is organized as follows. In Sec.\ref{anyonic_fermions} we briefly review the two-body color code model and emerging anyonic fermions. In Sec.\ref{encoding} we use these emerging excitations for encoding and implementation of single and two-qubit gates. The conclusions are presented in Sec.\ref{conclusion}.
\begin{figure}
\begin{center}
\includegraphics[width=8cm]{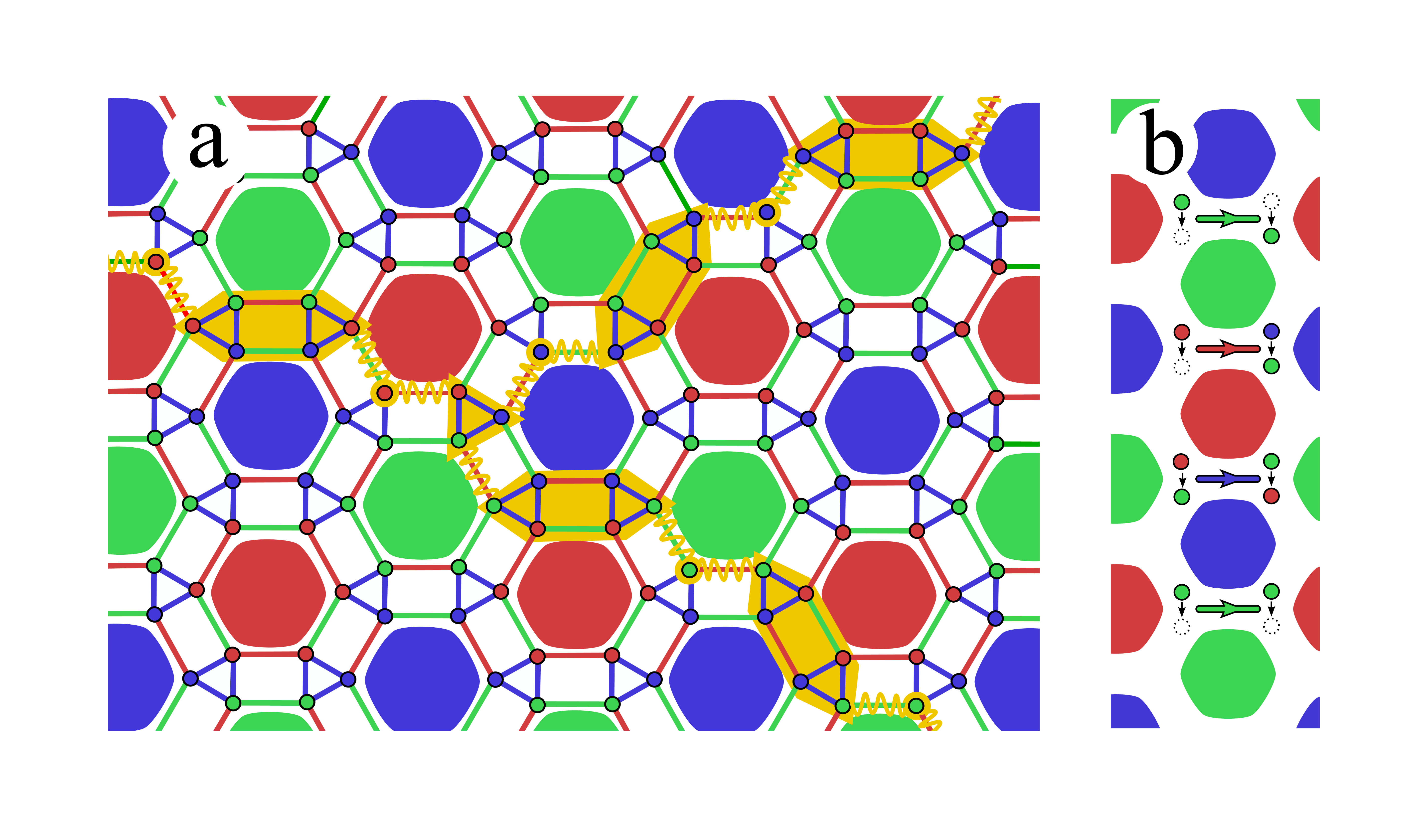}
\caption{(color online) (a) A piece of rubby lattice. Links are colored according
to the interaction between spins sitting at the vertices. Blue, red
and green links stand for $\sigma^{z}\sigma^{z}$,
$\sigma^{x}\sigma^{x}$ and $\sigma^{y}\sigma^{y}$, respectively. The
plaquettes can also be colored accordingly, and the spins around a
plaquette have the sam color with plaquette. A string-net is also shown, where  three strings with different colors
can meet at an effective site. (b) A pictorial representation of terms in Eq.\ref{Terms}. From up to bottom: 
hopping, fusion, color switching and annihilation.}
\label{fig1}
\end{center}
\end{figure}
\section{emerging anyonic fermions}\label{anyonic_fermions}
To see how high energy fermions appear, consider a set of
spin-$1/2$ particles sitting at the vertices of the rubby lattice as shown in
Fig.\ref{fig1}. We assume the following interactions hold between spins.
\bea \label{eq1}
H=-J_{x}\sum_{r-link}\sigma^{x}_{i}\sigma^{x}_{j}-J_{y}\sum_{g-link}\sigma^{y}_{i}\sigma^{y}_{j}-J_{z}\sum_{b-link}\sigma^{z}_{i}\sigma^{z}_{j}, \nonumber \\ \eea
where we are using three colors \emph{red}, \emph{green} and
\emph{blue} to distinguish between different links. Each
colored link represents an interaction as in Eq.\ref{eq1}. The
plaquettes of the lattice are colored and vertices can also be
colored accordingly. Associated with each colored plaquette, say blue hexagon in Fig.\ref{fig1}, three local string operators can be realized.\cite{kargarian1,kargarian2} Such local operators, called plaquette operators, commute with each other and with the Hamiltonian in Eq.\ref{eq1}. Hence, they are constants of motion and can be used to identify the local symmetry of the model. Let's denote them by $P_1$, $P_2$ and $P_3$. The explicit expression of these operators are given in Appendix.\ref{plaquette_operators}. By use of Pauli algebra, it's immediate to check both of the following relations.
\bea \label{plaquette} P_{1}^2=P_{2}^2=P_{3}^2=1,~~~P_1P_2P_3=-1.\eea
These relations indicate that all three plaquette operators are not independent giving rise to the local gauge symmetry $\mathrm{Z_2}\times \mathrm{Z_2}$ of the color code model. 

For any path on the lattice one can realize
string operators in which the contribution of the vertices to the
corresponding operator is determined by the outgoing link at each
vertex. For example if a vertex on the string is crossed by a red link, its contribution to string operator is $\sigma^{x}$. A simple example of elementary string operators is shown in Appendix. \ref{plaquette_operators}. Strings can be combined to form string-nets.\cite{kargarian1,kargarian2} A typical string-net is shown in Fig.\ref{fig1}, where each colored string
connects plaquettes with same color and three strings meet at a
triangle. The number of integrals of motions is exponentially increasing. Let $3N_s$ be the total number of
spins, so the number of plaquettes will be $N_s/2$. Regarding to the gauge symmetry of the model,
the number of independent plaquette operators is $N_s$. This implies that there are $2N_s$ integrals of motion and allow us to divide the Hilbert space into $2^{N_s}$ sectors labeled by the eigenvalues of plaquette operators. However, the Hamiltonian Eq.\ref{eq1} can not be exactly solved based on the integrals of motions. Moreover, because of four-valent structure of the lattice, unlike the Kitaev model the model
is not exactly solvable in terms of mapping to Majorana
fermions.\cite{kargarian1,feng:prl07} However, many features of the physics of the model can be
treated by considering the limiting behavior of the model.

We divide the Hamiltonian in two parts as $H=H_0+V$, where the first term denotes the unperturbed Hamiltonian $H_0=-J_{z}\sum_{b-link}\sigma^{z}_{i}\sigma^{z}_{j}$ and the second term denotes the interaction between triangles as $V=-J_{x}\sum_{r-link}\sigma^{x}_{i}\sigma^{x}_{j}-J_{y}\sum_{g-link}\sigma^{y}_{i}\sigma^{y}_{j}$. In the isolated triangle limit, i.e. $J_{x}=J_{y}=0$, the lattice
contains a collection of blue triangles. For each triangle the
subspace spanned by polarized spins, up or down, has the minimum
energy of $-3J_{z}$. Thus, the ground state of a triangle is
two-fold degenerate and the excited states with energy of $J_{z}$
are six-fold degenerate. So the ground state of Hamiltonian is
highly degenerated spanned by different configuration of polarized
spins on triangles.

As the couplings $J_{x}$ and $J_{y}$ grow on, the ground state
degeneracy is broken. The effects of perturbations can be teated by
invoking the spin-boson transformation.\cite{kargarian1,hector3} The transformation is exact
in which the spectrum of each triangle is replaced by an effective
spin and an hard-core boson. Taking all possible
degrees freedom of spins on a triangle into account, we find four possible cases
for a boson to be on the triangle: nothing, red, green and blue. In fact the excitation of each triangle is revealed by a colored boson. The mapping from original degrees of freedoms to effective spin and hard-core bosons is given in Appendix.\ref{appx_mapping}. From now on triangles are addressed by effective sites forming a hexagonal lattice $\Lambda$. Thus, we distinguish between sites (triangles) and vertices. The letter $c$ is used for one of the colors, then a bar
operation $\bar{c}$ transforms colors cyclically as $\bar{r}=g$,
$\bar{g}=b$ and $\bar{b}=r$. We also use the notation convention $c|c:=x$,$\bar{c}|c:=y$ and $\bar{\bar{c}}|c:=z$.

We refer to a site by considering its position relative to a reference site: the notation $O_{,c}$ means $O$ applied at the site that is connected to a site of reference by a link with color $c$.
The different terms in the Hamiltonian can then be interpreted in terms of effective
spins and hard-core bosons as follows.\cite{kargarian1,hector3}
\bea \label{Hamiltonian_B}
H=-3N/4+Q-\sum_{\Lambda}\sum_{c\neq c^{\prime}} J_{c^{\prime}|c}T_c^{c^{\prime}}, \eea
with $N$ the number of sites ($N=N_s/3$),
$Q:=\sum_\Lambda n$ the total number of hardcore bosons, the first
sum running over the $N$ sites of the hexagonal lattice, the second
sum running over the 6 combinations of different colors $c,c^{\prime}$
and \bea \label{Terms}
T_c^{c^{\prime}} = u_c^{c^{\prime}}+ \frac {t_c^{c^{\prime}}+v_c^{c^{\prime}}} 2 + \frac {r_c^{c^{\prime}}} 4 + \mathrm{h.c.}, \eea
a sum of several terms for an implicit reference site, according to
the notation convention defined above. The explicit expression of the terms appearing in the above equation is given in Appendix.\ref{appx_mapping}. They represent
various bosonic processes including $t_c^{c^{\prime}}$: hopping bosons between sites, $v_c^{c^{\prime}}$: annihilation or creation of a pair of bosons, $u_c^{c^{\prime}}$: fusion of two bosons in two another
one or splitting of one boson into two others and $r_c^{c^{\prime}}$: switching between
colors of two bosons.\cite{hector3} A pictorial representation of these terms is shown in Fig.\ref{fig1}b. We will use these processes as
possible ways for encoding of qubit states. 

By examining the hopping terms\cite{wen} it is simple to see that a
colored hard-core boson is fermion, i.e. under the exchange of two
hard-core bosons with same color a minus sign arises. This implies
that in this model we are dealing with three classes of high-energy
fermions each of one color. Although hard-core bosons in each class are fermions by
themselves, they have mutual semionic statistics with respect to fermions in 
other classes. This means that if for example a blue high-energy
fermion go around a green high-energy fermion, the wavefunction picks up a minus sign. This is the reason for the name of anyonic
fermions.\cite{hector3}

The appearance of such anyonic fermions is in sharp contrast with
emerging high-energy fermions in the Kitaev
model.\cite{vidal1,vidal2} In the latter model we are dealt with
one class of high-energy fermions. This is rooted in the fact that in the
Kitaev model there is only one type of strings carrying fermions,
while in the color code model there are three different classes of
strings forming string-net structure for the model. As long as the
anyonic properties of excitations are important as in implementing
one- and two-qubit gates, we can use colored high-energy fermions. This
is the subject of the next section. 
\section{encoding qubits and implementation of gates}\label{encoding}
As usual let $|0\rangle$ and $|1\rangle$ stand for the states of a
qubit in the computational bases. These states first must be encoded
in the states of the anyons, then manipulation of them is done
by anyons. The contribution of the colors in the
construction of the code makes it possible to use different methods
for encoding. Once the encoding is done, a corresponding
manipulation of anyons are assigned to implement gates.
\subsection{Encoding in hopping process}
\begin{figure}
\begin{center}
\includegraphics[width=8cm]{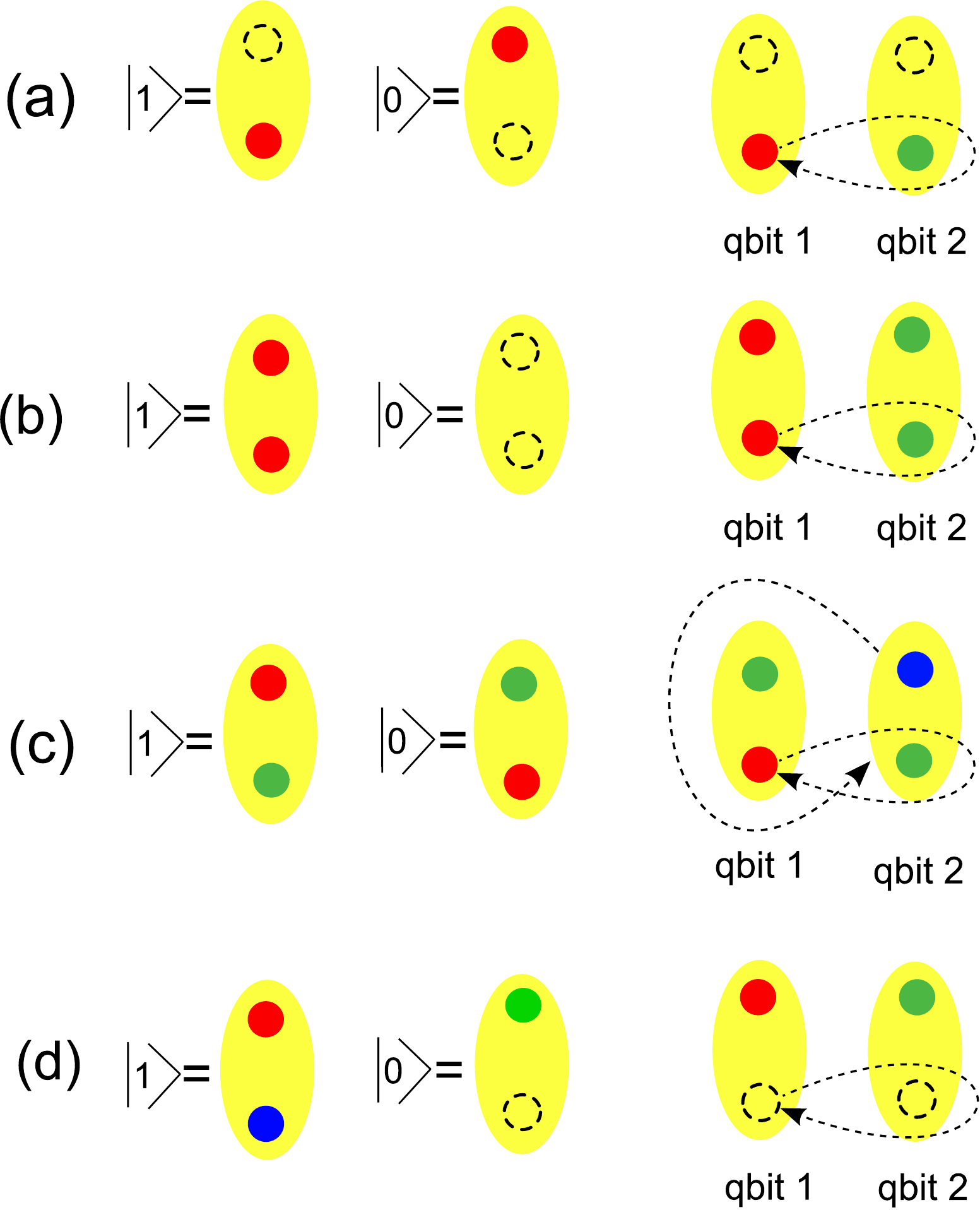}
\caption{(color online) Different processes that make the encoding
of a logical qubit (left) and the corresponding braiding process for
implementation of two-qubit gate (right). Each ellipse shows a two-site model, and arrow dashed line
indicates the braiding path. The processes are (a) the hopping process
(b) the annihilation process (c) the color switching process and (d)
the fusion process.} \label{fig2}
\end{center}
\end{figure}

The hopping of a $c$-fermion from a site to another one provides a
natural way for encoding the states of a qubit. To this end, we
encode the states of a qubit on a two-site model with an
$c$-fermion. We define a local Hilbert space with two bases as $|c,0\rangle$ and $|0,c\rangle$ if first and
second sites, respectively, are occupied by $c$-fermion. These
states can be used to encode two states of a qubit. If $c$-fermion
is on the first site, i.e. $|c,0\rangle$ , it corresponds to
$|1\rangle$. Otherwise, it will encode the $|0\rangle$. A pictorial
representation of such encoding has been shown in Fig.\ref{fig2}a.
Let $t_{c}$ operator stands for the hopping of $c$-fermion from one
site to another one. Thus, the hopping operator just encodes the
action of Pauli operator $\sigma^{x}$ in the space spanned by the
states of qubit. In fact, the hopping process is just the NOT gate.
The number operator $n_{c}$ at first site, which measures number of
$c$-fermions at a site, can be used to simulate the $\sigma^{z}$ in
the qubit space. In fact, the parity operator $p_{c}=1-2n_{c}$ gives
a minus (plus) sign if the first site is occupied (unoccupied). With
this realization of Pauli operators in terms of $c$-fermion process,
we have all required ingredients for implementation of single-qubit
$X$ and $Z$ gates. Notice that the latter gates are no longer performed in a topological 
way as we must switch between different states by means of local interactions, which are not 
protected. 

How can we implement two-qubit gates? As we will see two-qubit gates
require braiding of particles around each other. First we use
two pairs of sites each encodes one qubit. Two qubits must be encoded
by fermions with different colors in order for braiding to be
nontrivial. To perform a two-qubit gate, one simply needs to take
whatever is in the first site of the first qubit (control qubit),
then move it around the first site of the second qubit (target
qubit), as shown in Fig.\ref{fig2}a. Trivially, if the first qubit
is in the state $|0\rangle$, which corresponds to the situation in
which the first site is empty, the braiding action leaves the second
qubit unchanged. In other words, the second qubit is not braided any more.
However, the nontrivial case will be achieved whenever both qubits are in
the state $|1\rangle$. Only in this case the first sites of both
qubits are occupied by the fermions with different colors, namely first
site of control qubit is occupied with a $c$-fermion and first site
of the target qubit is occupied with a $\bar{c}$-fermion. Thus, the
braiding of $c$-fermion around the $\bar{c}$-fermion gives rise to
overall factor (here minus) for the states of two qubits. 

All possible cases of braiding can be summarized as follows: \bea
\nonumber &&|0,r\rangle|0,g\rangle\rightarrow
|0,r\rangle|0,g\rangle~~\Leftrightarrow~~|00\rangle\rightarrow
|00\rangle, \nonumber \\ &&|0,r\rangle|g,0\rangle\rightarrow
|0,r\rangle|g,0\rangle~~\Leftrightarrow~~|01\rangle\rightarrow |01\rangle, \nonumber \\
&&|r,0\rangle|0,g\rangle\rightarrow
|r,0\rangle|0,g\rangle~~\Leftrightarrow~~|10\rangle\rightarrow
|10\rangle, \nonumber \\ &&|r,0\rangle|g,0\rangle\rightarrow
-|r,0\rangle|g,0\rangle\Leftrightarrow~|11\rangle\rightarrow
-|11\rangle.\eea But the above evolution of states of two qubits
under braiding is just the controlled phase gate, which can be
turned into the CNOT-gate by applying the $\sigma^{x}$ to the target
qubit. The braiding process can be applied between any two arbitrary
qubits. We should only note that two qubits are encoded by fermions with different
colors, since fermions with different colors have
semionic mutual statistics. The braiding evolution is resilient to
any deformation of the braiding path so long as only the
first site of the target qubit is braided. The ability to
perform two-qubit gate together with one-qubit $X$ and $Z$ gates make the
implementation of part of a universal set of gates for quantum computation. For full implementation 
of universal gates, we should be able to implement two other single qubit gates, the Hadamard and phase gates, as well, which it seems that would not be possible in this setting.  
\subsection{Encoding by annihilation/creation process}
Again a two-site model is used to encode a single
qubit. An annihilation/creation process will annihilate/create a
pair of $c$-fermions on two sites. Let operator $v_{c}$ stands for
annihilation of pair of $c$-fermions on two sites. Given the empty
or filled sites, the states of a qubit can be simply encoded. The
empty sites, i.e. $|0,0\rangle$, and filled sites, i.e.
$|c,c\rangle$, are properly adjusted to encode the states
$|0\rangle$ and $|1\rangle$ of qubit, respectively. A schematic
representation of this encoding is shown in Fig.\ref{fig2}b.
The annihilation and creation operators switch between two different
states of two-site model, which in the logical space is interpreted
as Pauli operator $\sigma^{x}$. The parity of either first or second
sites can be used to encode the $\sigma^{z}$ Pauli operator in the
logical space. Therefore, the annihilation/creation and parity
operators allows one to perform single-qubit $X$ and $Z$ gates in the logical
space of a qubit.

Implementation of two-qubit gate is done by braiding of
$c$-fermions. Two qubits are encoded in two pairs of sites
separately. But the corresponding logical space of each qubit
carries a distinct color. For example, as shown in
Fig.\ref{fig2}b, we assume the logical space of first qubit
(control) has red fermion while the second qubit (target) has green
fermion. Choosing fermions with different colors is important for using the benefit of 
nontrivial braiding between them. We suppose a process through which
the contents of the first site of control qubit move around the first
site of target qubit. Definitely, the logical states $|00\rangle$
and $|01\rangle$ remains unchanged through braiding since the first
site of target qubit is not braided at all. The logical state
$|10\rangle$ also doesn't change since the $c$-fermion moves around
an empty site. However, when the $\bar{c}$-fermion of target qubit
is braided by the $c$-fermion of control qubit, a global phase (here
minus sign) arises. This latter case corresponds to the evolution of
logical state $|11\rangle$ into $-|11\rangle$ that is just the
controlled-phase gate. 

Regarding to Fig.\ref{fig2}b, the resulted
two-qubit gate is as follows \bea \nonumber
&&|0,0\rangle|0,0\rangle\rightarrow|0,0\rangle|0,0\rangle
~~\Leftrightarrow~~ |00\rangle\rightarrow|00\rangle, \\ \nonumber
&&|0,0\rangle|g,g\rangle\rightarrow|0,0\rangle|g,g\rangle
~~\Leftrightarrow~~ |01\rangle\rightarrow|01\rangle, \\ \nonumber
\nonumber&&|r,r\rangle|0,0\rangle\rightarrow|r,r\rangle|0,0\rangle~~\Leftrightarrow~~
|10\rangle\rightarrow|10\rangle, \\
&&|r,r\rangle|g,g\rangle\rightarrow-|r,r\rangle|g,g\rangle\Leftrightarrow~
|11\rangle\rightarrow-|11\rangle. \eea Therefore, the encoding of
logical states into annihilation/creation process and braiding
evolution provide an alternative way for constructing a set
of gates.

\subsection{Encoding in color switching process}
Two colored fermions with different colors can interchange their
colors by tuning interaction between spins as in
Hamiltonian of Eq.\ref{eq1}. To have a concrete discussion
consider two sites carrying fermions each of one color. A pictorial
representation is shown in Fig.\ref{fig2}c, where we used red and
green colors for fermions. Let $|r,g\rangle$ be a state of fermions
in which first and second sites carry red and green fermions,
respectively. An color switching operator $r_{c}=r_{c,1}r_{c,2}$,
where $r_{c,1}$ and $r_{c,2}$ act on first and second sites,
respectively, takes the state $|\bar{c},\bar{\bar{c}}\rangle$ and
transforms it as
$r_{c}|\bar{c},\bar{\bar{c}}\rangle=|\bar{\bar{c}},\bar{c}\rangle$.
Thus for each color $c$ two states $|\bar{c},\bar{\bar{c}}\rangle$ and
$|\bar{\bar{c}},\bar{c}\rangle $ are used to encode the
computational states of a logical qubit. The respective qubit is
referred as $c$-qubit. When a switching operator turns on, one may
take it into account as the operation of $\sigma^{x}$ in the
computational bases. Thus the transition from $|0\rangle$ into
$|1\rangle$ can be performed by means of color switching. A phase
shift of a qubit is operated by parity measurement of the first
site. Defining $p_{\bar{\bar{c}}}=1-2n_{\bar{\bar{c}}}$, it leaves
$|\bar{c},\bar{\bar{c}}\rangle$ unchanged, but gives a minus sign
upon the measurement of $|\bar{\bar{c}},\bar{c}\rangle$ as
$p_{\bar{\bar{c}}}|\bar{\bar{c}},\bar{c}\rangle=-|\bar{\bar{c}},\bar{c}\rangle$.
It results in rising phase difference between $|0\rangle$ and
$|1\rangle$ states.

As before, the realization of two-qubit gates is based on the braiding of
$c$-fermions sitting at sites. A two-qubit
gate must be implemented between two qubits with different colors. Namely,
we should consider $c-$ and $\bar{c}-$qubits. For instance, one may
consider the situation depicted in Fig.\ref{fig2}c, where we are
dealing with a $b$-qubit as qbit1 and $r$-qubit as qbit2. The
states of either qubit are encoded according to their colors as
above. In particular, as shown in Fig.\ref{fig2}c, the states of the
encoded control qubit are $|r,g\rangle$ and $|g,r\rangle$, and the
corresponding states of the target qubit are $|g,b\rangle$ and
$|b,g\rangle$. Now we can offer an instruction for the braiding of
fermions that permits the implementation of two-qubit gate. The
instruction includes two sequential braiding processes: (i) the
first site of control qubit moves around the first site of target qubit,
 then (ii) the target qubit entirely moves around the control qubit
or vice versa. If the order of two processes are reversed,
the final result of braiding remains unchanged. Such processes are
shown in Fig.\ref{fig2}c. 

It is simple to check the effect of
above evolution on all possible states of two qubits. The results
are as follows. \bea \nonumber
&&|r,g\rangle|g,b\rangle\rightarrow|r,g\rangle|g,b\rangle
~~\Leftrightarrow~~ |00\rangle\rightarrow|00\rangle, \\ \nonumber
&&|r,g\rangle|b,g\rangle\rightarrow|r,g\rangle|b,g\rangle
~~\Leftrightarrow~~ |01\rangle\rightarrow|01\rangle, \\ \nonumber
\nonumber&&|g,r\rangle|g,b\rangle\rightarrow-|g,r\rangle|g,b\rangle\Leftrightarrow~
|10\rangle\rightarrow-|10\rangle, \\
&&|g,r\rangle|b,g\rangle\rightarrow|g,r\rangle|b,g\rangle~~\Leftrightarrow~~
|11\rangle\rightarrow|11\rangle, \eea which clearly manifest the
braiding process encodes the controlled-phase gate in the logical
space of two qubits. It is not the only way of performing two-qubit
gate. One may consider another scenario for braiding. First take the
first site of control qubit and move it around the fist site of
target qubit, then the first (second) site of either control or
target qubit is braided by second (first) site. They eventually give
rise to the above results of performing two-qubit gate in the
logical space. Being able to perform the braiding of
fermions around each other by either above methods, we see that the
color switching between fermions together with the braiding provide
what we need to perform the single qubit $X$ and $Z$ gates and two-qubit gate. 

\subsection{Encoding in fusion process}
A significant feature of the two-body color code mode is the 
existence of 3-vertex interaction in the interacting fermionic processes.\cite{hector3} The
existence of 3-vertex interaction exhibits in the representation in terms of effective sites, three colored fermions can fuse into vacuum making the high-energy
fermions highly interacting. In other words two high-energy fermions with
different colors can fuse to the third one. This is called a fusion
process. Let $u_{c}$ denotes the fusion operator that fuses a
$c$-fermion and $\bar{c}$($\bar{\bar{c}}$)-fermion into a
$\bar{\bar{c}}$($\bar{c}$)-fermion.

One may wonder if fusion process could be used to encodes the states
of a logical qubit. Again consider a two-site model. Suppose first
and second sites are occupied by blue and green fermions,
respectively, as shown in Fig.\ref{fig2}d. The result of fusion
process is that the first site is unoccupied and the the second site
is occupied by a red fermion. Indeed, through the fusion process the
fermion on the first site is annihilated and the color of fermion on
the second site is switched.  Let $|c,\bar{c}\rangle$ and
$|0,\bar{\bar{c}}\rangle$ stand for states of two-site model through
a fusion process. We exploit these two states to encode the
computational bases of a qubit. We call such a qubit as a
$\bar{\bar{c}}$-qubit. By definition, the state
$|0,\bar{\bar{c}}\rangle$ is used to encode $|0\rangle$ and the
state $|c,\bar{c}\rangle$ encodes $|1\rangle$. An imaginative
picture of this encoding is shown in Fig.\ref{fig2}d. The
fusion operator and its conjugate (splitting) transform the states
into each other, which eventually can be used to encode the
$\sigma^{x}$ Pauli operator in logical space. The parity operator
can be adapted in which encodes the $\sigma^{z}$. In fact the parity
of fist site leaves the $|0,\bar{\bar{c}}\rangle$ unchanged while
shifts the phase for the state $|c,\bar{c}\rangle$. Thus, the fusion
process has all necessary ingredients for implementation of $X$ and $Z$ gates.

As before performing two-qubit operation requires braiding of
fermions around each other. To have a nontrivial braiding process,
we need to encode two qubits in different sets of states
characterized by different colors. To have a concrete discussion,
let consider the case shown in Fig.\ref{fig2}d, where two qubits
(qbit1 and qbit2) are red and green ones. The red qubit serves
as control qubit and the green qubit serves as target one. With the
above definition for the states of the sites, the states that encode
the red qubit (qbit1) are as $|0,r\rangle$ and $|g,b\rangle$, and
the states that encode the green qubit (qbit2) are as
$|0,g\rangle$ and $|b,r\rangle$. The braiding scheme is simple:
the first site of the target qubit must be braided by the entire
contents of the first site of the control qubit, as shown in
Fig.\ref{fig2}d. If the control qubit is the state $|0\rangle$,
its first site carries no fermion resulting in a trivial braiding.
The same holds when the target qubit is in the state $|0\rangle$. By
inspection we see that the only nontrivial case happens when both
qubits are in the $|1\rangle$ state. The effect of braiding on all possible states of two
qubits are as follows. \bea \nonumber
&&|0,r\rangle|0,g\rangle\rightarrow|0,r\rangle|0,g\rangle
~~\Leftrightarrow~~ |00\rangle\rightarrow|00\rangle, \\ \nonumber
&&|0,r\rangle|b,r\rangle\rightarrow|0,r\rangle|b,r\rangle
~~\Leftrightarrow~~ |01\rangle\rightarrow|01\rangle, \\ \nonumber
\nonumber&&|g,b\rangle|0,g\rangle\rightarrow|g,b\rangle|0,g\rangle~~\Leftrightarrow~~
|10\rangle\rightarrow|10\rangle, \\
&&|g,b\rangle|b,r\rangle\rightarrow-|g,b\rangle|b,r\rangle\Leftrightarrow~
|11\rangle\rightarrow-|11\rangle. \eea Thus, we see that the
fusion process encodes the a single qubit and if supplemented with
braiding as stated above, the implementation of $X$ and $Z$ gates and CNOT becomes possible.
\section{summary and concluding remarks}\label{conclusion}
In this work the problem of implementation of a set of
gates, which includes single-qubit (X and Z) and two-qubit (CNOT) gates, using the
anyonic excitations of the two-body color code model was studied.
Because of its underlying lattice structure, we can realize three
types of strings each of one color that can form a string-net
structure. The excitation spectrum of the model can be realized by the existence of three families of high-energy excitations. Each family is characterized by a color charge. The particle-like excitation in
each family are fermions by themselves, but excitations from
different families have a semionic mutual statistics, i.e. they are
the anyonic fermions. This latter point is in sharp contrast with
Kitaev honeycomb model, where the high-energy excitations have
only fermionic statistics since they emerge from only one type of
string. Therefore, it seems that the anyonic properties of high-energy fermions in the
color code can be used for implementation of gates in quantum information tasks.

The fermions in the color code model undergo several processes such
as hopping from site to another, annihilation/creation of pair of
fermions on two sites, switching colors between fermions on two
sites and fusion of fermions. All of them are driven by the terms in
the Hamiltonian. Each process can be adjusted in such a way that
encodes the states of a logical qubit. Namely, the computational
bases are encoded in a two-site model including colored fermions,
and the respective operation of fermionic process is translated into
the Pauli operators. Thus, the $X$ and $Z$ single-qubit gates can be implemented by
fermionic process. Neither encoding qubits nor implementation of latter single-qubit gates are topologically 
protected, thus they may suffer from the local perturbations. However, a successful implementation depends on ability of controlling fermionic quasiparticles. 

Implementation of two-qubit gates require
braiding of fermions. The states of control and target qubits must
be encoded in separate pairs of sites. One may build those pairs
in which they carry fermions with different color charges. In that case since
they belong to different families of fermions, the braiding process
can give rise to a nontrivial phase being suitable for performing of a two-qubit
gate, the CNOT gate. Once a qubit is encoded in a fermionic process,
the corresponding braiding encodes the controlled-phase gate. The
implementation is topological as the path of braiding does not
matter so long as an colored fermion is braided by another one.

Although performing the single-qubit gates are not topological, it
is done in a way that is different from the manipulation of anyons
in the Kitaev model. In the latter model implementation of
single-qubit gates is done in a dynamical way, which needs rotation of spins to alter the anyonic state leading to
switching between computational bases. But in the color code model
the manipulation of colored fermions via any processes is driven by
interaction terms in the Hamiltonian. Therefore, one only needs to control
the interacting terms in which the proper manipulation occurs. Another advantage of our 
construction is that a quantum state of hard-core boson can be simulated in optical lattices.\cite{exp_hardcore} 

One may think that the process of braiding could accumulate 
errors in the system. But by a method based on the trapping
potential,\cite{pachos} the particle can move around immune. 
Therefore, the fermionic processes in the color code model encode the
logical qubit and due to the nontrivial braiding of fermions with 
different color charge, the implementation of some gates
becomes possible. An arbitrary quantum information task includes
preparation of an initial state, implementation set of universal
gates and measurements. The initial state can be prepared by an
array of two-site models and filling the sites with colored fermion. 
The implementation of single-qubit gates ($X$ and $Z$) are performed by fermionic
process introduced in preceding section and the two-qubit gates are
done through the proper braiding of fermions around each other.
Finally, the resulted state can be measured by determining how sites
have been occupied by fermions.

\acknowledgements
We gratefully acknowledge financial support from ARO Grant W911NF-09-1-0527 and NSF Grant DMR-0955778. We also acknowledge the hospitality of Institute For Research In Fundamental Sciences (Tehran, Iran) where this work was initiated.

\appendix
\section{Bosonic mapping}
\label{appx_mapping}
The mapping between the original spin degrees of freedom on a triangle and effective spin coupled to a hard-core boson is as follows.\cite{kargarian1,hector3}
\bea \nonumber
|\!\Uparrow,0\rangle\equiv|\!\uparrow\uparrow\uparrow\rangle,~~~~~&&|\!\Downarrow,0\rangle\equiv|\!\downarrow\downarrow\downarrow\rangle\\
\nonumber |\!\Uparrow,\mathrm{r}\rangle\equiv|\!\uparrow\downarrow\downarrow\rangle,~~~~~&&|\!\Downarrow,\mathrm{r}\rangle\equiv|\!\downarrow\uparrow\uparrow\rangle \\
\nonumber |\!\Uparrow,\mathrm{g}\rangle\equiv|\!\downarrow\uparrow\downarrow\rangle,~~~~~&&|\!\Downarrow,\mathrm{g}\rangle\equiv|\!\uparrow\downarrow\uparrow\rangle\\
|\!\Uparrow,\mathrm{b}\rangle\equiv|\!\downarrow\downarrow\uparrow\rangle,
~~~~~&&|\!\Downarrow,\mathrm{b}\rangle\equiv|\!\uparrow\uparrow\downarrow\rangle,
\label{mapping_bases} \eea 
where the $\Uparrow$($\Downarrow$) and $\uparrow$($\downarrow$) stand for the states of effective spin sitting at a site and original spin sitting at a vertex of rubby lattice, respectively. 

At each site we can
introduce the color annihilation operator as $b_{c} :=
|0\rangle\langle c|$. The number operator $n$ and color number
operator $n_{c}$ are \be n :=\sum_c n_c, \qquad n_c := b^\dagger_c
b_c. \ee
In terms of operators, the mapping Eq.\ref{mapping_bases} can be
expressed as follows.
\begin{equation}\label{mapping} \sigma^z_c \equiv \tau^z \otimes
p_c, \quad \sigma^\nu_c \equiv \tau^\nu \otimes (b_c^\dagger + b_c +
s_\nu r_c), \end{equation} where $\nu=x,y$, $s_x:=-s_y:=1$, the
symbols $\tau$ denote the Pauli operators on the effective spin and
we are using the color parity $p_c$ and the $r_c$ operators defined as
\begin{equation}\label{parity}
p_c := 1-2(n_{\bar c}+n_{\bar{\bar{c}}}),\qquad r_c :=
b^\dagger_{\bar{c}} b_{\bar{\bar{c}}} + b^\dagger_{\bar{\bar{c}}}
b_{\bar{c}}.\end{equation} 
 
The explicit expressions of terms appearing in Eq.\ref{Terms} are
\bea
&t_c^{c'} := \tau_c^{c'} b_c b^\dagger_{c, c'},~~ r_c^{c'} :=
\tau_c^{c'} r_c r_{c, c'},\nonumber \\ &u_c^{c'} :=  s_{c'|c} \tau_c^{c'} b_c
r_{c, c'},~~ v_c^{c'}:= \tau_c^{c'} b_c b_{c, c'}, \label{terms}
\eea where we are using the notation
\begin{equation}
\tau_c^{c'}:=\tau^{c'|c}\tau_{, c'}^{c'|c}.
\end{equation}
\begin{figure}
\begin{center}
\includegraphics[width=8cm]{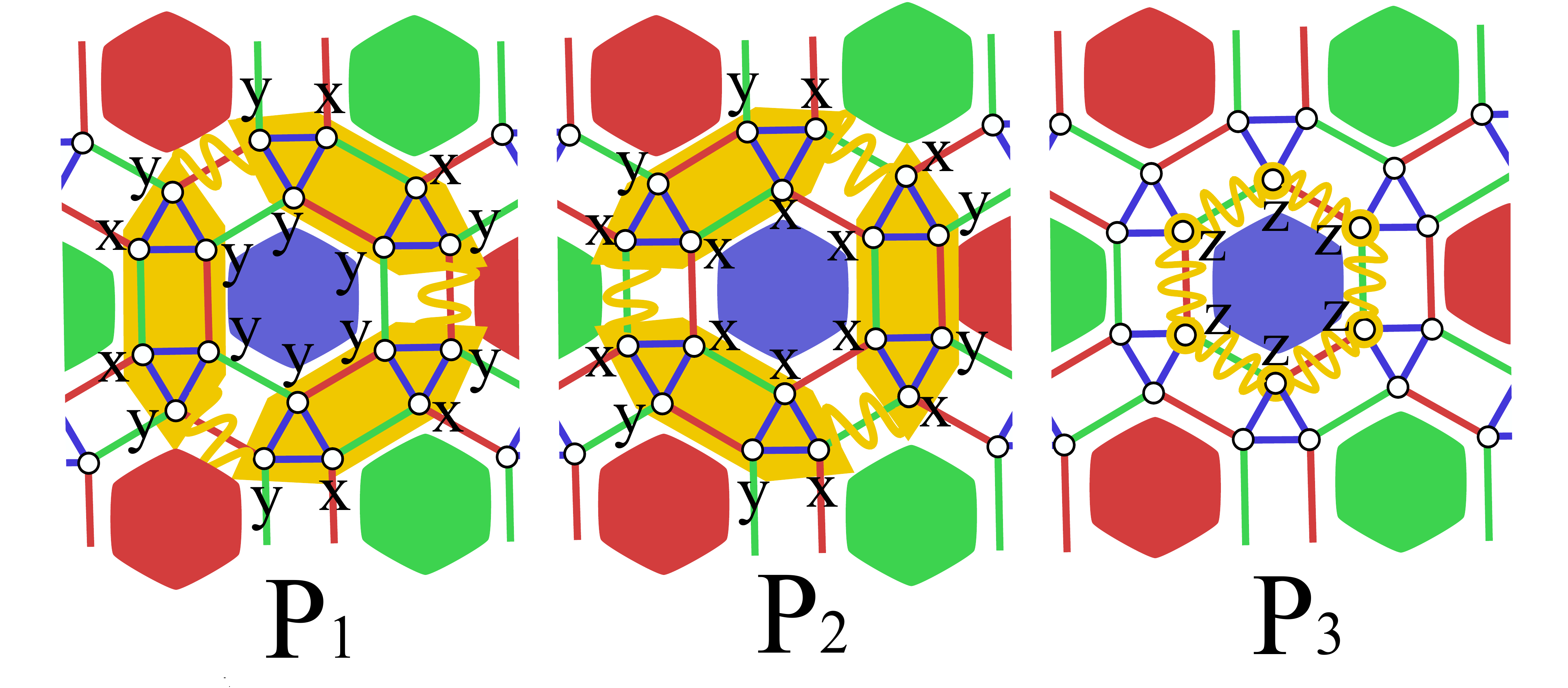}
\caption{(color online) A representation of an elementary plaquette. With each plaquette we attach three local operators shown by yellow (light) string. The explicit expression of these operators is given by product of Pauli matrices acting on vertices labeled by $x$, $y$ and $z$ standing for $\sigma^{x}$, $\sigma^{y}$ and $\sigma^{z}$, respectively.}
\label{fig3}
\end{center}
\end{figure}

\section{Plaquette operators}\label{plaquette_operators}
On the rubby lattice a plaquette is defined by an inner hexagon with six triangles surrounding it. So each plaquette has eighteen vertices as shown in Fig.\ref{fig3}. With each plaquette we attached three plaquette operators $P_1$, $P_2$ and $P_3$ having following properties.
\bea [P_i,P_j]=0~~\mathrm{and}~~[P_i,H]=0~~\forall i,j.\eea The explicit expressions are as follows. \bea P_1=\prod_{l} \sigma^{\nu_l}_{l},~P_2=\prod_{k} \sigma^{\nu_k}_{k},~P_3=\prod_{m} \sigma_{m}^{z},\eea where products in $P_1$ and $P_2$ go over eighteen vertices with Pauli matrices $\sigma^{x}$ and $\sigma^{y}$, and product in $P_3$ goes over six vertices with all $\sigma^{z}$ as shown in Fig.\ref{fig3}. $\nu_l$ and $\nu_k$ stand for $x$ or $y$ as shown besides vertices. Note that such identifications for local strings can be used to color the plaquettes and vertices of the lattice. We define the $P_1$ string as red string in which its fat parts (stretched yellow hexagon) connect red plaquettes, and we use green color for $P_2$. The third string is defined as a blue string surrounding around the internal hexagon. With this convention other plaquettes of lattice are colored accordingly. Having colored hexagons, we can simply color vertices: they have same color with the inner hexagon.     


\section*{References}
\begin{thebibliography}{99}

\bibitem{nielsen}
M. Nielsen and I. Chuang, \emph{Quantum Computation and Quantum
Information}, Cambridge University Press, 2000.

\bibitem{miguel:rmp02}
A. Galindo and  M. A. Martin-Delgado, 
Rev. Mod. Phys. {\bf 74}, 347 (2002). 

\bibitem{jean}
Jean-Luc Brylinski, Ranee Brylinski, \emph{Mathematics of Quantum Computation}, Chapman-Hall/CRC Press, 2002. quant-ph/0108062.

\bibitem{mochon}
C. Mochon, Phys. Rev. A \textbf{67}, 022315 (2003).

\bibitem{freedman}
M. Freedman, M. Larsen, and Z. Wang, Comm. Math. Phys. \textbf{227},
605(2002).

\bibitem{nayak:rmp08}
Chetan Nayak, Steven H. Simon, Ady Stern, Michael Freedman and Sankar Das Sarma, 
Rev. Mod. Phys. {\bf 80}, 1083 (2008).

\bibitem{hector:arxiv09}
Hector Bombin, Ravindra W. Chhajlany, Michal Horodecki, and Miguel-Angel Martin-Delgado, arXiv:0907.5228.

\bibitem{hamma:arxiv09}
Alioscia Hamma, Claudio Castelnovo, Claudio Chamon, Phys. Rev. B \textbf{79}, 245122 (2009)

\bibitem{dennis}
E. Dennis, A. Kitaev, A. Landahl, and J. Preskill, J. Math. Phys.
(N.Y.) \textbf{43}, 4452 (2002).

\bibitem{kitaev1}
A.Yu. Kitaev, Ann. Phys. (N.Y.) 303, \textbf{2} (2003); \emph{ibid}
2, \textbf{321} (2006).

\bibitem{Lloyd}
S. Lloyd, Quant. Inf. Proc. 1, 13 (2002). 

\bibitem{pachos}
J. K. Pachos, Int. J. Quant. Info, \textbf{4}, No.6 (2006) 947.

\bibitem{vidal1}
K.P. Schmidt, S. Dusuel, J. Vidal, Phys. Rev. Lett. \textbf{100},
177204 (2008).

\bibitem{vidal2}
J. Vidal, K.P. Schmidt, S. Dusuel, Phys. Rev. B \textbf{78}, 245121
(2008).

\bibitem{hector1}
H. Bombin and M. A. Martin-Delgado, Phys. Rev. Lett. \textbf{97},
180501 (2006).

\bibitem{hector2}
H. Bombin and M. A. Martin-Delgado, Phys. Rev. A. \textbf{76},
012305 (2007).

\bibitem{kargarian1}
M. Kargarian, H. Bombin and M. A. Martin-Delgado, New J. Phys. \textbf{12}, 025018 (2010). 

\bibitem{kargarian2}
 H. Bombin, M. Kargarian and M. A. Martin-Delgado, Fortsch. Phys. \textbf{57}, 1103 (2010).

\bibitem{hector3}
H. Bombin, M. Kargarian and M. A. Martin-Delgado, Phys. Rev. B \textbf{80}, 075111 (2009).

\bibitem{expriment1}
Xing-Can Yao, Tian-Xiong Wang, Hao-Ze Chen, Wei-Bo Gao, Austin G. Fowler, Robert Raussendorf, Zeng-Bing Chen, Nai-Le Liu, Chao-Yang Lu, You-Jin Deng, Yu-Ao Chen, Jian-Wei Pan, Nature \textbf{482}, 489-494 (2012).

\bibitem{expriment2}
Hendrik Weimer, Markus M\"uller, Igor Lesanovsky, Peter Zoller, Hans Peter B\"uchler, Nature Phys. \textbf{6}, 382 (2010).

\bibitem{exp_hardcore}
B. Capogrosso-Sansone, C. Trefzger, M. Lewenstein, P. Zoller, G. Pupillo, Phys. Rev. Lett. \textbf{104}, 125301 (2010).

\bibitem{feng:prl07}
X.-Y. Feng, G.-M. Zhang, T. Xiang,  Phys. Rev. Lett. \textbf{98}, 087204 (2007).

\bibitem{wen}
M. Levin, X.-G. Wen, Phys. Rev. B \textbf{67}, 245316 (2003).

\end{thebibliography}

\end{document}